\documentclass[10pt]{iopart}
\usepackage{epsfig}

\def\lesssim{\mathrel{\hbox{\rlap{\hbox{\lower4pt\hbox{$\sim$}}}\hbox{$<$}}}}
\def\gtrsim{\mathrel{\hbox{\rlap{\hbox{\lower4pt\hbox{$\sim$}}}\hbox{$>$}}}}
\newcommand{\tf}{t_{\mathrm{f}}}
\newcommand{\Msun}{\mathrm{M}_{\odot}}
\newcommand{\dy}{\,\mathrm{days}}

\begin{document}

\title[Identifying Supermassive Black Hole Binary Inspirals]{Identifying Decaying Supermassive Black Hole Binaries from their Variable Electromagnetic Emission}

\author{Zolt\'an Haiman$^1$, Bence Kocsis$^{2,3}$, Kristen Menou$^1$, Zolt\'an Lippai$^4$, and Zsolt Frei$^4$}

\address{$^1$ Department of Astronomy, Columbia University, New York, USA\\
$^2$ Harvard-Smithsonian Center for Astrophysics, Cambridge, USA\\
$^3$ School of Natural Sciences, Institute for Advanced Study, Princeton, NJ, USA\\
$^4$ Institute of Physics, E\"otv\"os University, Budapest, Hungary}

\begin{abstract}
Supermassive black hole binaries (SMBHBs) with masses in the mass
range $\sim (10^4$--$10^7)\,{\rm M_\odot}/(1+z)$, produced in galaxy
mergers, are thought to complete their coalescence due to the emission
of gravitational waves (GWs).  The anticipated detection of the GWs
by the future {\it Laser Interferometric Space Antenna} ({\it LISA})
will constitute a milestone for fundamental physics and astrophysics.
While the GW signatures themselves will provide a treasure trove of
information, if the source can be securely identified in
electromagnetic (EM) bands, this would open up entirely new scientific
opportunities, to probe fundamental physics, astrophysics, and
cosmology.
We discuss several ideas, involving wide--field telescopes, that may
be useful in locating electromagnetic counterparts to SMBHBs detected
by {\it LISA}. In particular, the binary may produce a variable
electromagnetic flux, such as a roughly periodic signal due to the
orbital motion prior to coalescence, or a prompt transient signal
caused by shocks in the circumbinary disk when the SMBHB recoils and
"shakes" the disk. We discuss whether these time-variable EM
signatures may be detectable, and how they can help in identifying a
unique counterpart within the localization errors provided by LISA.
We also discuss a possibility of identifying a population of
coalescing SMBHBs statistically, in a deep optical survey for
periodically variable sources, before {\it LISA} detects the GWs
directly.  The discovery of such sources would confirm that gas is
present in the vicinity and is being perturbed by the SMBHB -- serving
as a proof of concept for eventually finding actual {\it LISA}
counterparts.
\end{abstract}



\section{Introduction}

The anticipated detection by {\it LISA} of gravitational waves emitted
during the coalescence of supermassive black holes (SMBHs) in the mass
range $\sim (10^4$--$10^7)\,{\rm M_\odot}/(1+z)$ will constitute a
milestone for fundamental physics and astrophysics.  While the GW
signatures themselves are a rich source of information, if the GW
source produces electromagnetic (EM) radiation, and if the object can
be securely identified in EM bands, this would open up entirely new
scientific opportunities. The simultaneous study of photons and
gravitons from a single source could probe fundamental aspects of
gravitational physics \cite{dm07, paper2}.  The GW sources can also be
used as self--calibrated standard sirens \cite{sch86}, and
cosmological parameters can be determined if the source redshift is
identified \cite{hh05,koc06}.  Finally, for many events in the above
mass and redshift range, {\it LISA} will be able to measure the masses
and spin vectors of the SMBHs, their orbital parameters, and their
luminosity distance, to a precision unprecedented in any other type of
astronomical observation \cite{lh06,vec04}.  If a counterpart is
known, then the Eddington ratios and other attributes of black hole
accretion physics can be studied in exquisite detail
\cite{koc06,paper2}.

Motivated by the above possibilities, several recent studies have
addressed the question of whether finding a counterpart will be
feasible, given {\it LISA}'s localization errors
\cite{hh05,koc06,paper1,lh07,paper2}.  The crucial uncertainty, of
course, is the nature (luminosity, spectrum, and time--evolution) of
any EM emission produced by coalescing SMBHBs during the GW--inspiral
stage.  Such emission would have to be related to gas in the vicinity,
and possibly accreting onto the coalescing SMBHs.  The gas around the
BHs would likely settle into a rotationally supported, circumbinary
disk.  In a geometrically thin disk, the torques from the binary
create a central cavity, nearly devoid of gas, within a region about
twice the orbital separation \cite{al94} (for a nearly equal--mass
binary), or a narrower gap around the orbit of the lower--mass BH in
the case of unequal masses $q \equiv M_1/M_2 \ll 1$ and larger orbital
separations \cite{an02}.  In the latter case, the lower--mass hole
could ``usher'' the gas inward as its orbit decays, producing a prompt
and luminous signal during coalescence.  In the former case, residual
gas flow into the cavity and onto the BHs, such as suggested in
numerical simulations \cite{al96,mm08,cuadra08}, may still produce
non--negligible EM emission.  Around the time of coalescence, the
gravitational waves shear the circumbinary gas and could brighten its
emission detectably~\cite{kl08}. Finally, SMBHBs recoil at the time of
their coalescence due to the emission of GWs, at speeds up to
4,000~${\rm km~s^{-1}}$ (e.g. ref.~\cite{campanelli07} and references
therein).  The gas disk will respond promptly (on the local orbital
timescale) to such a kick, which may produce prompt shocks, and
transient EM emission after coalescence~\cite{lippai}.\footnote{Such
brightening due to kick-induced disk heating can persist on much
longer timescales, and could produce detectable emission -- leading to
recent proposals that this can help identify a population of such
sources before {\it LISA}'s launch. This possibility will be briefly
discussed further in \S~\ref{sec:periodic} below.}

The complex processes involved in ultimately producing any EM emission
remain poorly understood, and the level of the luminosity produced, as
well as its spectrum and time-evolution, are essentially unknown. {\it
However, any emission during and promptly following the inspiral stage
is likely to be variable.}  In this contribution, we discuss three
issues related to using variability to identify EM counterparts. Can
LISA events be localized to within the field of view of astronomical
instruments (several deg$^2$), hours to weeks {\it prior} to
coalescence (\S~\ref{sec:trigger})?  What is the response of the
circumbinary gas to the gravitational recoil (``kick'') of the SMBHB
at coalescence (\S~\ref{sec:kick})?  Can we identify coalescing SMBHBs
before the launch of {\it LISA}, as variable sources, due to periodic
perturbations in the circumbinary gas (\S~\ref{sec:periodic})?

\section{Monitoring the 3D LISA Error Box Prior To Coalescence}
\label{sec:trigger}

The first and most fundamental question in searching for any EM
counterpart by looking for variable emission, during the final stages
of coalescence, is the accuracy to which the {\it LISA} source can be
localized at various look--back times {\em prior to the coalescence}.
(Coalescence is taken to occur when the binary separation reaches the
innermost stable circular orbit; ISCO).  Here we present
time--dependent localization errors, obtained by the Harmonic Mode
Decomposition method \cite{paper1}.  This technique uses the
restricted, post--Newtonian approximation for the GW waveform, and
applies the Fisher matrix technique to the Fourier transform of the
waveform, to forecast parameter uncertainties.  Orbits are assumed to
be circular, and spins are neglected (this is justified until the last
day or so of the merger; see \cite{lh07}). The 17--dimensional
parameter space describing the general binary inspiral is split into
``slow'' and ``fast'' parameters, based on the time--scales on which
they modulate the waveform, and the two sets are assumed to be
decoupled.  The angular coordinates of the source, and its luminosity
distance -- representing the 3D localization of the source -- are
``slow'' Fisher parameters, and information on these parameters are
derived from the annual motion of {\it LISA} around the Sun.  The
reader is referred to \cite{paper1} for a full list of assumptions and
details about the method.

\begin{figure}[t]
\begin{center}
  \includegraphics[height=0.235\textheight]{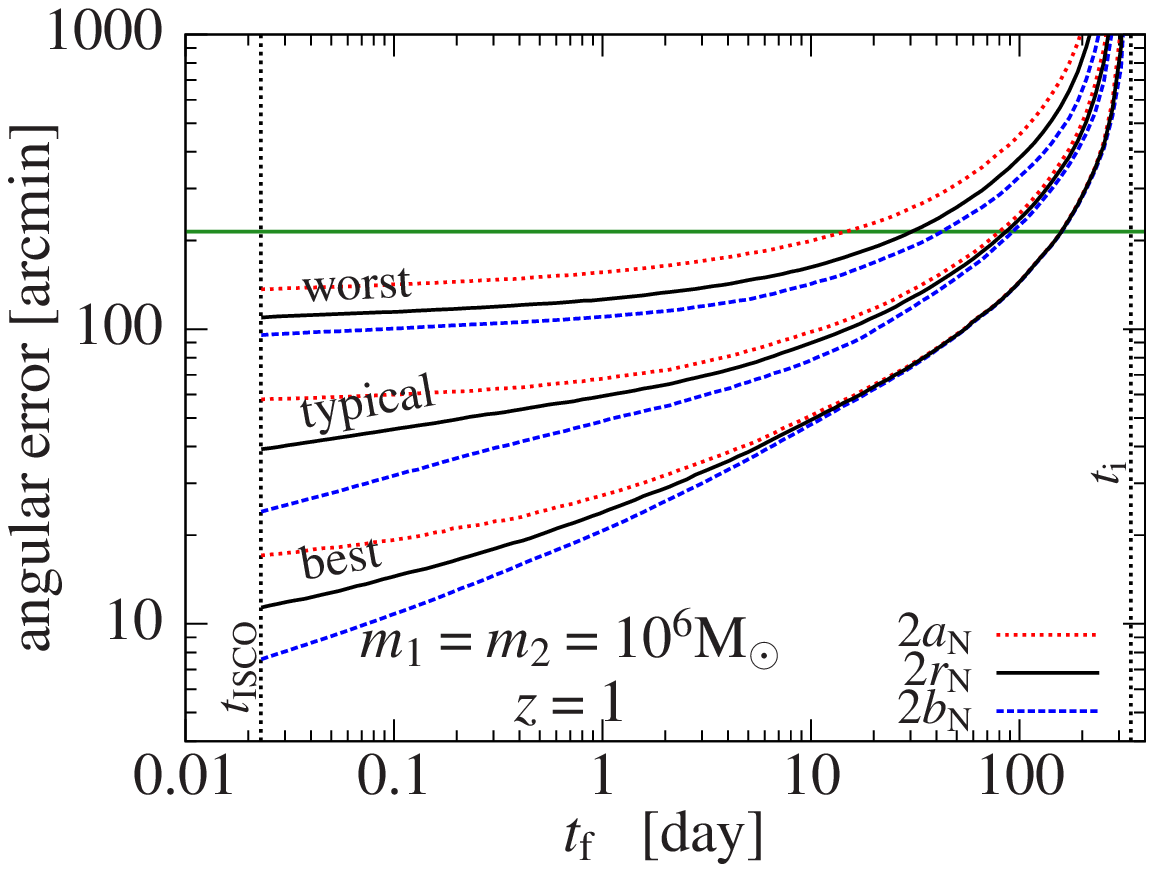}
  \includegraphics[height=0.235\textheight]{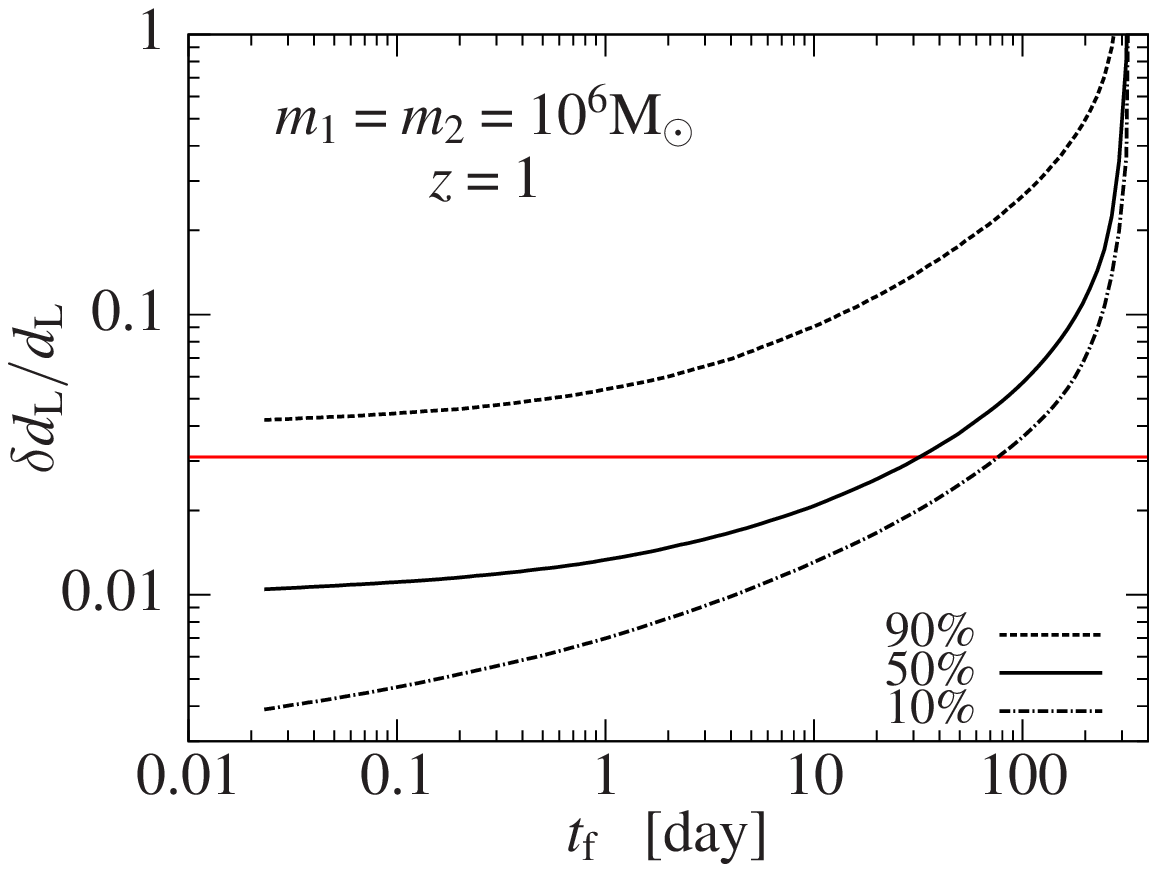}
\end{center}
\caption{\label{f:errortau} 
   Evolution with pre-ISCO look--back time,
  $\tf$, of the expected LISA source localization uncertainties.
  {\it Left Panel:} Sky position errors (major axis $2a_N$, minor axis
  $2b_N$, and equivalent diameter, $2r_N\equiv 2\sqrt{a_Nb_N}$, of the error
  ellipsoid). An equal--mass binary is shown as an example, with
  $M_1=M_2= 10^{6}\Msun$ at $z=1$.  Best, typical, and worst cases (among
  sources with random orientation) represent the $10\%$, $50\%$, and $90\%$
  levels of cumulative error distributions.  
  The absolute errors are typically small enough to fit
  in the FOV of a wide--field instrument in the last few weeks,
  allowing a world--wide monitoring campaign. The horizontal line
  shows a diameter of $3.57^{\circ}$, which corresponds to localizing
  the source to within 10 deg$^2$.
  {\it Right Panel:} Luminosity distance $d_L(z)$ errors for the same
  binary ($10\%$, $50\%$, and $90\%$ levels of cumulative error
  distributions as in the left panel).  The horizontal line delineates
  the level of weak lensing uncertainties (adopted from
  ref.~\cite{koc06}).  Knowing the luminosity distance to the
  accuracy of a few percent will allow the counterpart search to focus
  on candidates in a narrow redshift slice.}
\vspace{-1\baselineskip}
\end{figure}

The time--evolving $1\sigma$ errors on the two--dimensional sky
position are shown for an equal--mass binary, $M_1=M_2=10^6~{\rm
M_\odot}$, at redshift $z=1$ in Figure~\ref{f:errortau} (corresponding
roughly to the optimal choice of mass/redshift combination; other
masses and redshifts yield poorer localization) . The HMD method, by
construction, approximates the sky position errors by ellipses.
Figure~\ref{f:errortau} shows the gradual improvement in localization,
in the form of the major axis ($2a_N$), minor axis ($2b_N$) and
equivalent diameter ($2r_N=\sqrt{4a_Nb_N}$).  Figure~\ref{f:errortau}
displays results for three separate cumulative probability
distribution levels, $90\%,50\%,10\%$, so that $10\%$ refers to the
best $10\%$ of all events, as sampled by the random distribution of
five angular parameters.  The evolution of errors scales steeply with
look--back time for $\tf\gtrsim 40\dy$. For smaller look--back times,
errors essentially stop improving in the ``worst'' ($90\%$ level)
case, improve with a relatively shallow slope for the ``typical''
($50\%$ level) case, and improve more steeply in the ``best'' case
($10\%$ level among the realizations of fiducial angular parameters).
Similar evolutionary trends are seen for the luminosity distance
$d_L(z)$ errors (shown in the right panel in Fig.~\ref{f:errortau}).

\begin{figure}[t]
  \includegraphics[height=0.238\textheight]{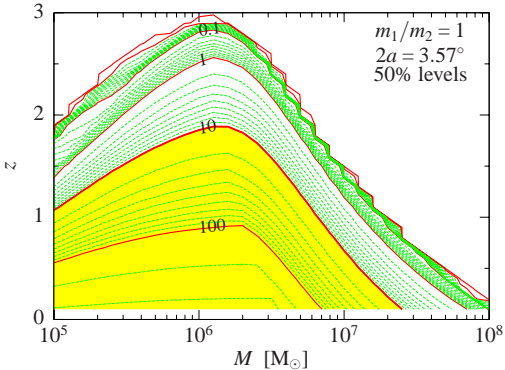}
\caption{\label{f:warning50} 
Contours of advance warning times in the total SMBHB mass ($M$)
vs. redshift ($z$) plane, for typical events with SMBH mass ratio
$q\equiv M_1/M_2=1$.  The contours trace the look--back times at which
the major axis ($2a$) of the localization error ellipsoid first falls
below $3.57^{\circ}$.  The contours are logarithmically spaced in
days, and $10$ days is highlighted with a thick curve. }
\vspace{-1\baselineskip}
\end{figure}

Figure~\ref{f:warning50} displays contours of fixed ``advance warning
time'' for typical ($50\%$) events, adopting a 10 deg$^2$ FOV as a
reference.  The contours are logarithmically spaced, with solid
contours every decade and the shaded region highlights the $(M,z)$
region where at least $10$--day advance notice will be available. This
figure shows that $10$ day advance warning to cover the full error
ellipsoid with a single LSST pointing is possible for a range of
masses and source redshifts, up to $M\sim 3\times 10^7\Msun$ and
$z\sim 1.7$.

The results shown in Figures~\ref{f:errortau} and \ref{f:warning50} 
suggest that it will be possible to identify, prior to merger, a small
enough region in the sky where any prompt electromagnetic (EM)
counterpart to a LISA inspiral event will be located. Given sufficient
advance notice, it will then be possible to trigger a world--wide
monitoring campaign, to search for EM counterparts as the merger
proceeds (and also during the most energetic coalescence phase).

The several square--degree field will, of course, contain a very large
number of sources (a few $\times10^5$ galaxies in total, at the
limiting optical magnitude of $\sim 27$ mag that may be relevant; e.g.
ref.~\cite{mad99} and discussion below).  Having predictions for the
spectrum and time--dependence of a coalescing SMBH binary, and
therefore knowing what to look for, would obviously greatly help in
identifying the counterpart, possibly allowing an identification even
post--merger.  Such a prediction, however, would require understanding
the complex hydrodynamics and radiative properties of circumbinary gas
at the relevant small separations of a few to a few thousand
Schwarzschild radii. This is a notoriously difficult problem even in
the much simpler case of steady accretion onto a single SMBH
\cite{mefa01}.  This suggests that the best strategy may be an
``open--minded'' search for any variable signatures prior and during
coalescence.

There will be several ways, however, to cut down on the list of
possible counterparts, using the photometric redshifts of the
candidates (restricting the search to within a narrow, $\delta z\lesssim$
few percent, redshift slice; see the right panel in
Fig.~\ref{f:errortau}), the expected luminosity of the source, and the
other parameters of the binary provided by {\it LISA} (for example, a
variable EM signal may be much more likely if the spin and the orbital
angular momenta are known to be aligned, as this may indicate the
presence of circumbinary gas).  These, and several other possible cuts
are discussed further in ref.~\cite{paper2}.

\section{Prompt Shocks in the Gas Disk Around a Recoiling Binary}
\label{sec:kick}

The recent break--through in numerical relativity has allowed a direct
computation of the linear momentum flux (``kick'') produced during the
coalescence of a SMBH binary.  Such kicks may help produce {\it
prompt} EM counterparts to GW sources detected by {\it LISA}.  If the
SMBHB is surrounded by a circumbinary gas disk, the disk will indeed
respond promptly (on the local orbital timescale) to such a kick. If
this results in warps or shocks, the disturbed disk could produce a
transient EM signature \cite{mp05}.  As discussed above, the final sky
localization uncertainty from {\it LISA} is typically a few tenths of
a square degree, containing a large number of sources; monitoring this
area for transient events {\it after} the merger may be another method
to securely identify counterparts.

We investigated the response of a circumbinary disk to the kick
\cite{lippai}.  We adopted the following simplified picture for the
disk around a fiducial, equal--mass, $M_1+M_2=10^6~{\rm M_\odot}$
binary.  The disk has an inner edge at $100 r_{\rm S}$ (Schwarzschild
radii), inside which it is empty (with the gas evacuated due to
torques from the binary; e.g. ref.~\cite{mp05}) and an outer edge at
$10,000 r_{\rm S}$. Outside of this radius, there may still be gas,
but it may be unstable to fragmentation, and it will, in any case,
evolve slowly (the behavior of this gas will then not be relevant for
a {\it LISA} counterpart search, but if emission is produced in this
gas, it could help identify SMBHBs independently; see discussion in
the next section).  The scale--height and temperature at the inner
edge is $h/r=0.46$ and $T=1.7\times10^6$K, respectively.  The
scale--height remains constant with radius out to $2,000 r_{\rm S}$,
beyond which it increases nearly linearly ($h\propto r^{21/20}$). The
temperature varies with radius as $T\propto r^{-9/10}$ (see
ref.~\cite{lippai} for further details).

The important features of such a disk (as well as other proposed
variants of thin $\alpha$-disks) are the following: (i) orbital
motions in thin disks are supersonic, so that the gas is susceptible
to shocks if disturbed; (ii) at the relevant radii outside $100r_{\rm
S}$ the viscous time--scale is long, and the orbits are near
Keplerian; (iii) gas near the inner edge of disk is tightly bound to
the kicked BHB ($v_{\rm orbit}\sim 2.1\times 10^4~{\rm km~s^{-1}}$),
but the outer edge ($v_{\rm orbit}\sim 2.1\times 10^3~{\rm
km~s^{-1}}$) can be marginally bound, or even unbound, for large kicks
(a rough condition for being bound is $v_{\rm orbit}\gtrsim v_{\rm
kick}$), and (iv) the total disk mass within $10,000 r_{\rm S}$ is
much less than the BHB mass, which justifies ignoring the inertia of
the gas bound to the BHB.

For a quantitative assessment of the disk's response to the kick, we
employed the following approximation: the disk particles are assumed
to be massless, collisionless, and initially on co-planar, circular
orbits. The kick simply adds the velocity $\vec{v}_{\rm kick}$ to the
instantaneous orbital velocity of each particle (in the inertial frame
centered on the BHB).  We used $N=10^6$ particles, distributed
randomly and uniformly along the two--dimensional surface of the disk.
The kick velocity was varied between $500~{\rm km~s^{-1}} < v_{\rm
kick} < 4,000~{\rm km~s^{-1}}$, and directed either perpendicular or
parallel to the initial disk plane.  Note that in both the
perpendicular and parallel case, we start with a two--dimensional
particle distribution (i.e. an infinitely thin disk), but in the
perpendicular case, we then follow the orbits in 3D.

\begin{figure}[t]
  \includegraphics[height=0.238\textheight]{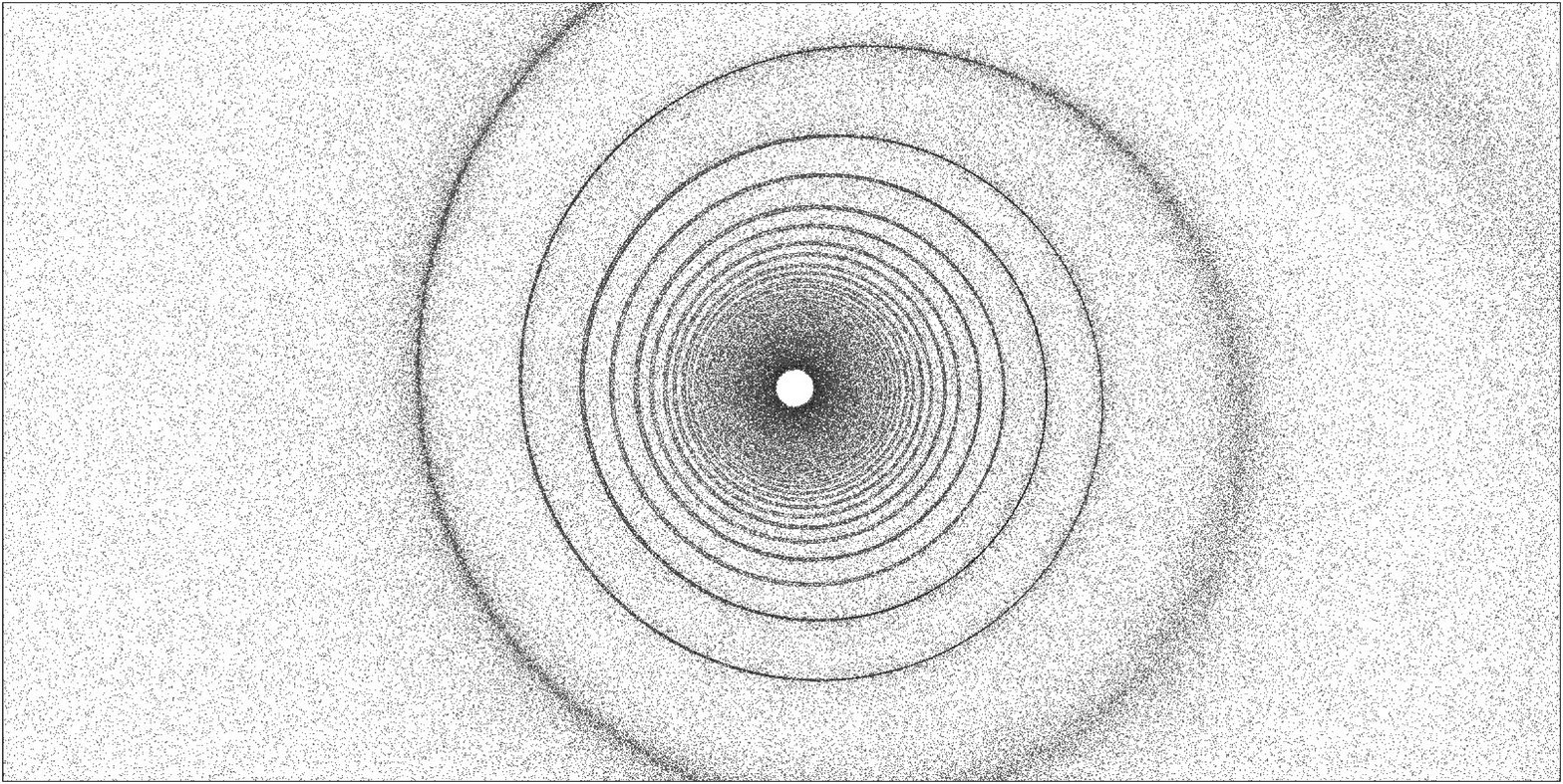}
  \includegraphics[height=0.238\textheight]{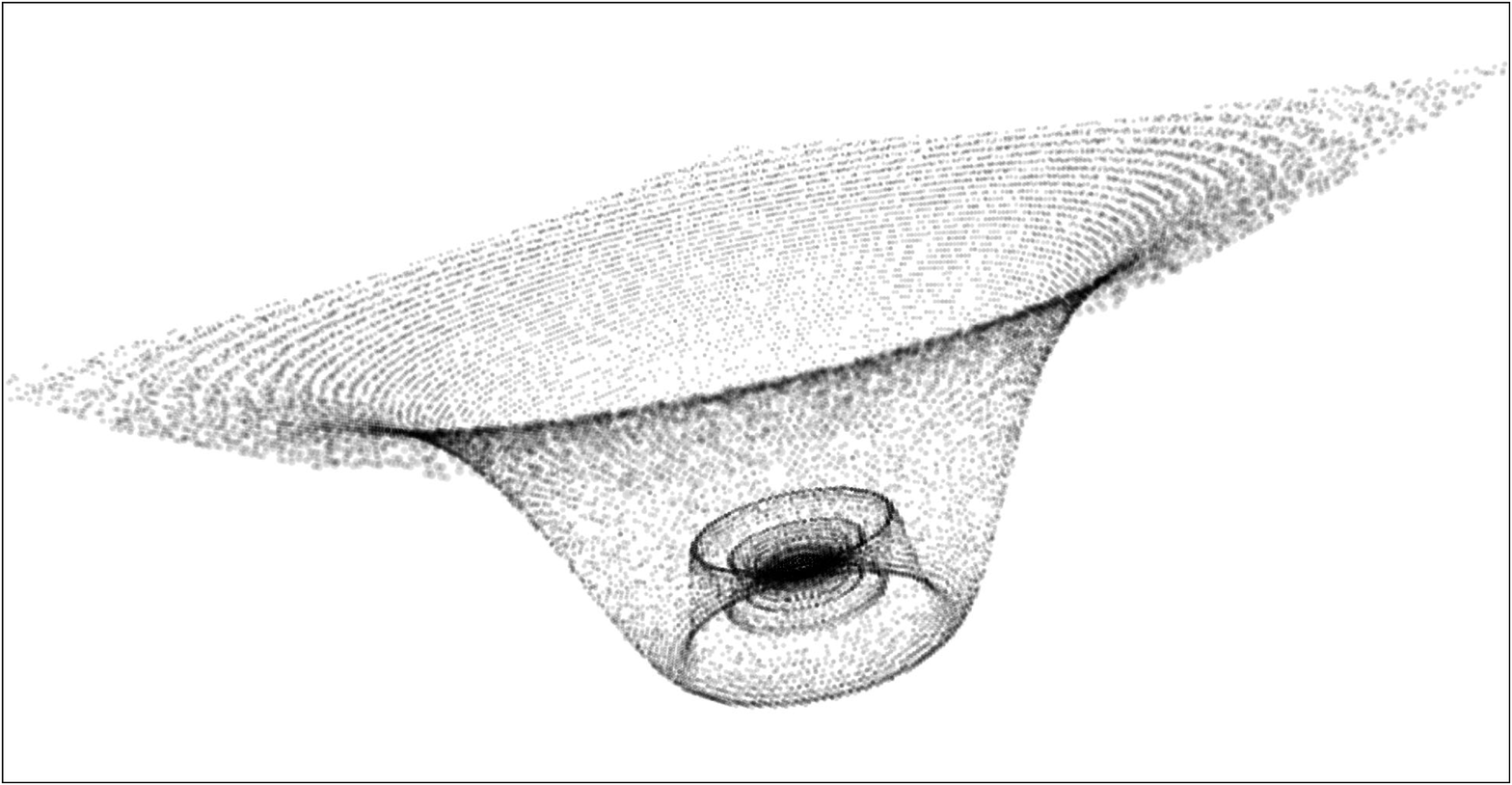}
\caption{\label{fig:kicks}
{\it Top Panel:} The top view of the surface density of a disk around
  an $M_1+M_2=10^6~{\rm M_\odot}$ BH binary, which recoiled within the
  disk plane at the velocity of $v_{\rm kick}=500~{\rm km~s^{-1}}$,
  oriented vertically upward in the diagram.  The disk is initially
  assumed to have an inner edge at $r_{\rm in}=100 r_{\rm S}$
  (Schwarzschild radii), and is shown here out to a radius of $r_{\rm
  in}=5,000 r_{\rm S}$, at a time $t=90$ days after the kick.  A
  tightly wound, outward--moving spiral caustic develops in about a
  month.  The dark/light shades correspond to regions of high/low
  density, the low--density regions being similar to the initial
  surface density, and the high--density regions about 10 times
  overdense.
{\it Bottom Panel:} Aerial view of a disk with a kicked SMBHB, as in
the left panel, except here the kick is oriented perpendicular to the
disk, and the snapshot is taken at $t=1$ week. For visual clarity, the
graphic contrast was increased relative to the left panel, and the
diagram stretched by a factor of 10 along the axis perpendicular to
the disk plane.  The density distribution is azimuthally symmetric,
and while there are mild concentric density fluctuations, strong
enhancements (i.e. caustics) were found to develop only after a delay
of $\approx$ one year.}
\vspace{-1\baselineskip}
\end{figure}

Figure~\ref{fig:kicks} shows, as an example, a face--on view of the
surface density of the disk $90$ days after a kick with $v_{\rm kick}
= 500~{\rm km~s^{-1}}$ in the plane of the disk (left panel).  The
sharp, tightly wound spiral features clearly seen in the figure trace
the locus of points where particles cross each other, corresponding
formally to a density caustic.  The spiral caustic first forms at
$\sim 30$ days, and then propagates outward at a speed of $\approx
500~{\rm km~s^{-1}}$, so that the outermost caustic at time $t_c$ is
located at $r_c\sim t_c v_{\rm kick}$ (this behavior can be roughly
understood using epicycles; refs.~\cite{lippai,sk08}). The right panel
in Figure~\ref{fig:kicks} shows, for comparison, the aerial view of
the 3D particle density one week after a kick with the same velocity,
but perpendicular to the disk.  The density profile in this case
remains azimuthally symmetric, but still develops concentric rings of
density fluctuations (although we find the density enhancements are
much weaker, at the ten percent level).

These results suggest that strong density enhancements can form
promptly after a supersonic kick in the plane of the circumbinary
disk, within a few weeks of the coalescence of a $\sim 10^6~{\rm
  M_\odot}$ BHB.  Because the disk is cold, and caustics are formed
when particles first cross each other along their orbits, this implies
that corresponding shocks could occur in a gas disk.  For
hydrodynamical shocks to occur within a finite--pressure gas, the
relative motions $v_{\rm c}$ between the neighboring particles that
produce the caustic must exceed the sound speed.  At the outermost
radius where the disk is marginally bound to the BHB, one expects
$v_{\rm c}\sim v_{\rm kick} \sim v_{\rm orbit}$; relative motions will
be slower further inside.  The relative speed should roughly
correspond to covering the epicyclic amplitude $\sim (v_{\rm
  kick}/v_{\rm orbit}) r_c$ in the caustic--formation time $t_c\sim
r_c/v_{\rm kick}$, yielding $v_c\sim v_{\rm kick}^2/v_{\rm orbit}$.
For $v_{\rm kick}= 500~{\rm km~s^{-1}}$, this predicts $v_{\rm c} \sim
25~{\rm km~s^{-1}} (r/1000r_{\rm S})^{1/2}$; we have verified in our
simulations that particles cross the caustics at speeds within
$\sim 30\%$ of this predicted value.  Compared with the sound speed
$c_s\approx 25~{\rm km~s^{-1}} (r/1000r_{\rm S})^{-9/20}$, this
suggests that the density waves produced by the kick in the gas beyond
$\sim 700r_{\rm S}$ will indeed steepen into shocks.  We also found
that the inclination of the kick may be important in determining the
strength and timing of such shocks.

The nature of the emission resulting from the shocks or density
enhancements will have to be addressed in future work, by
incorporating the effects of hydrodynamics, computing the heating rate
at the location of the spiral shocks, and modeling the overall disk
structure and vertical radiation transport.\footnote{Our preliminary
work, based on hydrodynamical simulations, does indicate that gas
disks with realistic pressure profiles still develop strong
shocks~\cite{lia08}.}  However, in the fiducial case discussed above,
if we assume that the shocked gas is heated to temperatures
corresponding to $v_{\rm c}$, and $t_{\rm c}$ is the time--scale on
which the corresponding thermal energy is converted to photons, then
we find that the luminosity may be a small but non--negligible
fraction, 0.2 percent to 5 percent of the Eddington luminosity of the
central BH, and would increase with time roughly as $L_{\rm kick}
\propto t^{2}$. This suggests that the afterglows may be detectable,
at least for nearby BHs and/or for the most massive BHs in LISA's
range (which extends up to $M_{\rm bh}\approx 10^7 {\rm M_\odot}$). We
may also speculate on the spectral evolution of the ``kick
after--glow'', based on the characteristic shock velocity at each
radius.  We find that the luminosity is dominated by the outermost
shocked shells, with the spectrum peaking at the characteristic photon
energy corresponding to $k T_{\rm shock}\propto v_{\rm c}^2 \propto
v_{\rm orbit}^{-2} \propto r$.  The shocks could therefore result in
an afterglow, starting from $700 r_{\rm S}/v_{\rm kick}\sim 50$ days,
first peaking in the UV band ($\sim 3$eV or $\sim 0.3\mu$m,
corresponding to $\sim 25~{\rm km s^{-1}}$), and then hardening to the
EUV/soft X--ray ($\sim 50$eV) range after $\sim$two years.  The
detection of such an afterglow would help identify EM counterparts to
{\it LISA} events.

\section{Searching for Gravitational Binary  Inspirals Before LISA}
\label{sec:periodic}

Numerical simulations suggest that the central cavity of the
circumbinary disk is not completely empty, and that there can be
non--negligible gas inflow into this cavity from the outside disk
\cite{al96,mm08,cuadra08}.  The perturbations of the circumbinary gas
by the rotating quadrupole potential of the binary, in fact, appears
to impose large fluctuations on this inflow rate, tracking the orbital
period \cite{mm08,cuadra08}.  We must emphasize that the level and
nature of EM emission, associated with this inflow (or with other
effects from the gas in the vicinity of the binary during the late
stages of binary evolution), remains essentially unknown.
Furthermore, even in the most optimistic scenario, in which a clearly
periodic emission is indeed produced, at a significant and detectable
flux level, there remains the (potentially severe) observational
challenge of distinguishing such variability from other possible
sources of periodic variations.

Despite these uncertainties and caveats, however, it is interesting to
ask the following question: if, indeed, non--negligible emission is
produced during the late stages of binary evolution, and the
luminosity varies periodically on the orbital time--scale, could such
periodic sources possibly be identified in EM surveys, even before
{\it LISA} becomes operational?

As the orbit of each individual binary shrinks, its orbital period,
and therefore, by assumption, its variability timescale decreases.
The evolution of a binary embedded in a thin circumbinary disk
generically proceeds through three distinct stages (see
ref.~\cite{periodic}).
{\it (i)} First, the binary is strongly coupled to the circumbinary
disk and is driven by viscosity (analogous to ``disk-dominated''
planetary migration), and the radius of the gap follows the binary.
As the binary separation shrinks below $\sim 10^5 r_{\rm S}$ ($10^3
r_{\rm S}$) for mass ratios $q\equiv M_1/M_2 \sim 1$ ($q\sim 0.01$),
the local disk density starts to dominate over the binary mass, and
the binary evolves more slowly, according to so--called
``secondary--dominated'' Type-II migration. During this stage, the GW
emission is negligible.
{\it (ii)} Later, within the radius $\sim 500 r_{\rm S}$, the binary
starts to be driven primarily by GWs but the radius of the gap can
still follow the binary.
{\it (iii)} Finally, within $\sim 100 r_{\rm S}$ the binary is
entirely driven by GWs and the binary falls in much more quickly than
the outer edge of the gap is able to move inward.
The ordering of these events is valid for a very broad range of binary
and disk parameters. Note that the outcome of the gas inside the
binary's orbit is left unspecified above (see ref.~\cite{an02} for a
possible outcome for $q\ll 1$).

\begin{figure}[t]
\begin{center}
  \includegraphics[height=0.312\textheight]{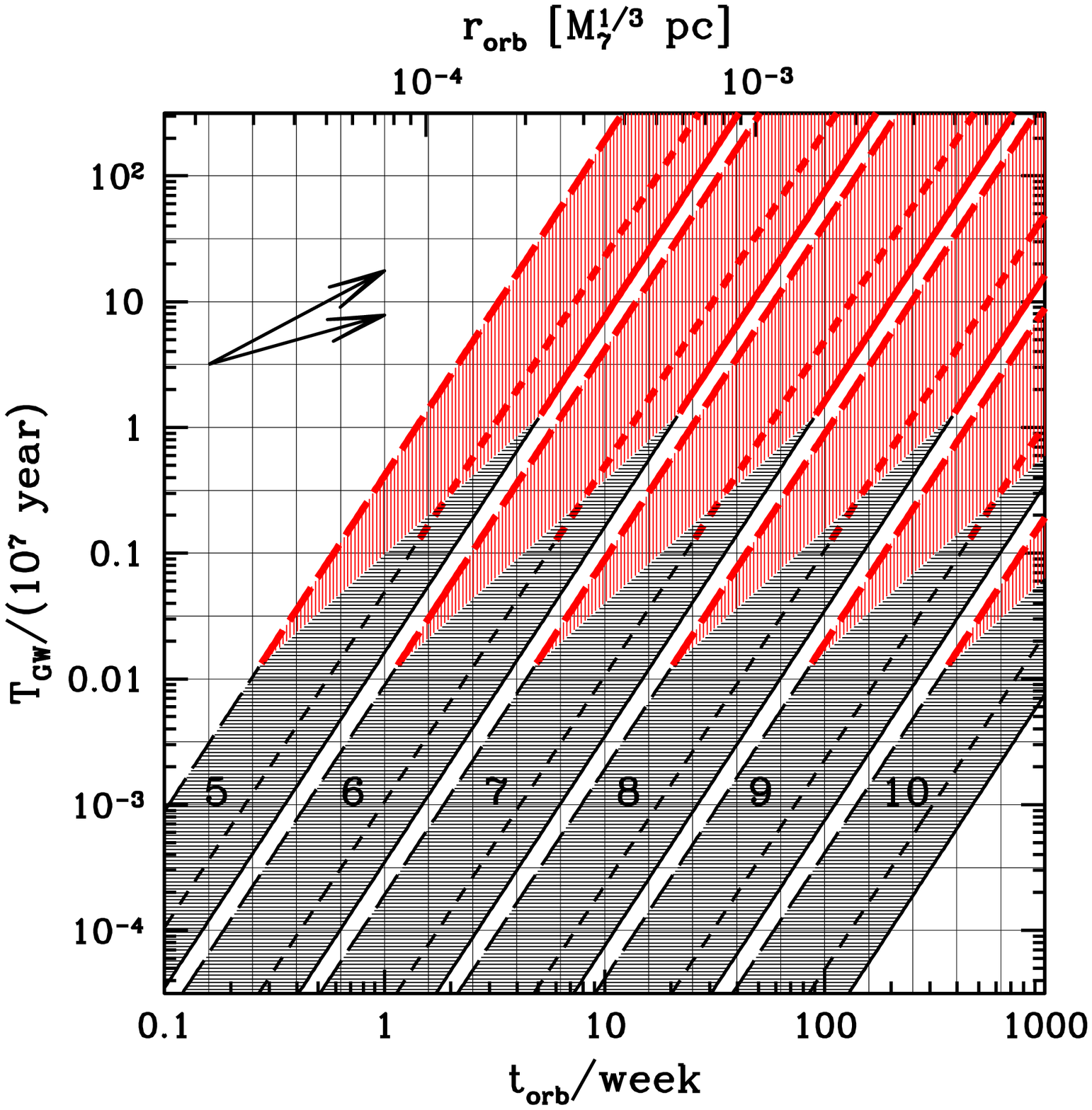}
  \includegraphics[height=0.312\textheight]{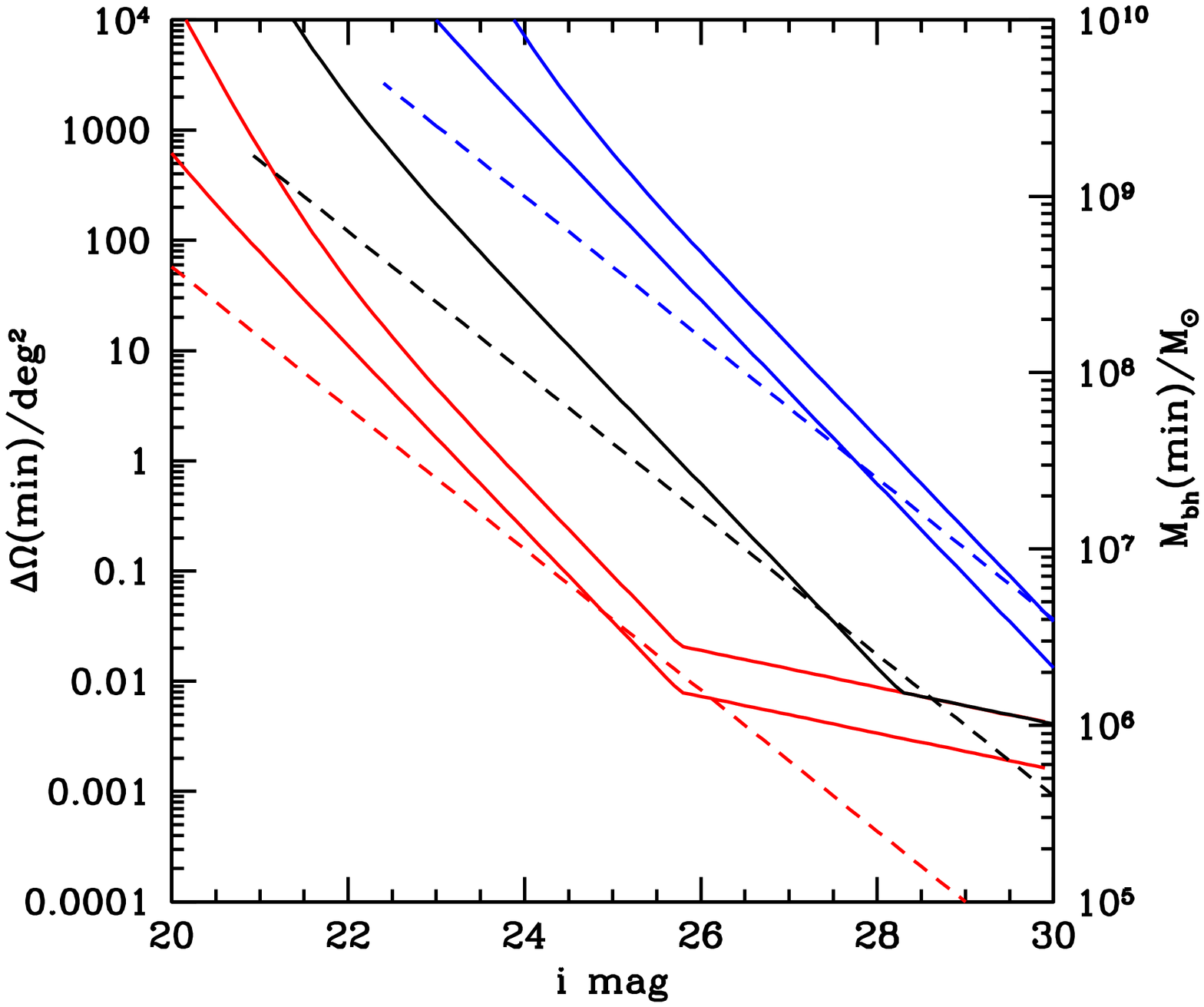}
\end{center}
\caption{\label{f:periodicsources} {\it Left Panel:} The figure shows
the time $T_{\rm GW}$ that a SMBHB, whose coalescence is driven
entirely by gravitational radiation, spends at a radius where the
orbital time is within $t_{\rm orb}$.  Both $T_{\rm GW}$ and $t_{\rm
orb}$ are in redshifted units (as measured on Earth).  Each of the
shaded regions corresponds to a different total BH mass ($[1+z]M_{\rm
tot}=[10^{5,6,7,8,9,10}]\Msun$; top to bottom), and shows three
different mass ratios ($q=[1, 0.1, 0.01]$; bottom to top, or solid,
short-dashed, and long-dashed, respectively).  The quantity on the $y$
axis can be interpreted as the duty--cycle for each source during
which it exhibits periodic variability on the time--scale $t_{\rm
var}\approx t_{\rm orb}$. The thick (red) portion of the lines and the
vertically shaded (red) areas indicate the orbital separations, for
each mass, where binary evolution may no longer be dominated by GWs.
The slope of the arrows indicate the flatter evolution of the binary
when it decays due to viscous torques: the flatter arrow corresponds
to the slope $T\propto t_{\rm orb}^{25/51}$ for near--equal mass
binaries, and the slightly steeper arrow to $T\propto t_{\rm
orb}^{14/15}$ for $q\ll 1$ (see Appendix in ref.~\cite{periodic} for
details). For reference, the $x$ axis labels on the top show the
orbital radius corresponding to each orbital time; for $M=10^7\Msun$
(or $M_7=1$) and $z=0$, the plotted range covers
$1.6\times10^{-5}-7.5\times10^{-3}{\rm pc}$, or 17-77,000
Schwarzschild radii.\\  \\

  {\it Right Panel:} This panel shows the sky coverage of a future
survey, required to find at least one periodic source, with a period
of at most 20 weeks, as a function of the $i$--band magnitude limit on
the variable component. This will ensure that a multi--year survey,
could find many more sources with longer periods, e.g. periods up to 1
year, demonstrating the periodic nature of the variations, and
allowing to deduce the abundance of periodic sources as a function of
period.  The three solid curves show results assuming that the
variable component is a fraction 0.3, 0.03, and 0.003 of the Eddington
luminosity of the binary.  Each dashed curve shows the corresponding
mass of the SMBHB of which there would be a single example of a
$t_{\rm var}=20$--week periodic variable in the survey ($\sim 20$
sources with the same luminosity are then predicted with a $t_{\rm
var}=60$ week period in the same survey volume, assuming pure
GW--driven evolution, or fewer such longer--period sources, if
viscosity speeds up the evolution significantly).}
\vspace{-1\baselineskip}
\end{figure}

In the last, GW--driven regime, the binary spends a characteristic
time $T_{\rm GW}$ at each orbital separation $r_{\rm orb}< r_{\rm GW}$
that scales with the corresponding orbital time as $T\propto t_{\rm
orb}^{8/3}$. The incidence rate of sources that have similar inferred
BH masses, and show near--periodic variability on the time-scale
$t_{\rm var}\approx t_{\rm orb}$, would then follow $N\propto t_{\rm
var}^{8/3}$.  On the other hand, at larger separations, where the
evolution is due to viscous processes, the incidence rate has a much
flatter dependence.  For near--equal mass binaries, the scaling is
between $N\propto t_{\rm var}^{25/51}$ and $N\propto t_{\rm
var}^{7/12}$, depending on whether the opacity is dominated free--free
absorption or electron scattering, respectively. For $q\ll 1$, the
scaling is slightly steeper, between $N\propto t_{\rm orb}^{5/6}$ and
$N\propto t_{\rm orb}^{14/15}$ for free--free absorption or electron
scattering opacity, respectively (see the Appendix in
ref.~\cite{periodic}).  In the left panel of
Figure~\ref{f:periodicsources}, we show that the time spent at each
orbital period can be interestingly long, i.e. a non--negligible
fraction of the expected lifetime of quasar activity of a $\sim$ few
$\times10^7$ years.

Can we detect the flux variations from these sources?  Luminosity
variations corresponding to a fraction $f_{\rm Edd}\lesssim 0.01$ of
the Eddington luminosity (or, e.g., a periodic component with
amplitude $i\approx 26+2.5\log[(f_{\rm Edd}/0.01) (M_{\rm BH}/3\times
10^7{\rm M_\odot})^{-1}]$ magnitudes in the optical for BHBs at $z=2$)
would have not been found in existing variability surveys. However, as
shown in Figure~\ref{f:periodicsources}, a dedicated, deep survey of a
$\sim$deg$^2$ area, possibly with existing instruments, could detect a
population of $\sim$several to $\sim$several thousand such
``transition sources'' with a range of periods between 20 weeks
$\lesssim t_{\rm var}\lesssim$ 60 weeks.  In the right panel of
Figure~\ref{f:periodicsources}, we use the luminosity function of
optical quasars, and assume that each optical quasar results from a
SMBH--SMBH merger, and, as the binary orbit decays, it goes through
the evolutionary stages discussed above (see ref.\cite{periodic} for
more details). The figure then shows the required depth and area
coverage for a multi--year survey to detect such periodic sources,
covering a factor of three range in period.  The dependence in the
abundance of these sources on their period -- in particular, a switch
from a flat, viscosity--driven power--law between $N\propto t_{\rm
var}^{1/2}$ and $N\propto t_{\rm var}^{1}$, to the much steeper,
GW--driven $N\propto t_{\rm var}^{8/3}$ -- below a characteristic
$t_{\rm var}\sim$ tens of weeks (corresponding to $\sim 10^3$
Schwarzschild radii for equal--mass $\sim 10^6 {\rm M_\odot}$
binaries), could confirm (1) that the orbital decay for sources below
a characteristic $t_{\rm var}$ is indeed driven by GWs and (ii) also
that circumbinary gas is present at small orbital radii and is being
perturbed by the BHs -- serving as a proof of concept for eventually
finding actual {\it LISA} counterparts.

Finally, we mention two other possibilities to identify coalescing
SMBHBs before {\it LISA}'s launch.  First, \cite{sb08} and \cite{sk08}
followed the response of the gas disk around a recoiling SMBHB,
similar to our calculation described in \S~\ref{sec:kick}, but on much
longer timescales ($\sim 10^4$ years).  Ref.~\cite{sb08} found that
the shocked gas, thrown out of the disk plane by an oblique kick, may
produce bright flares in X-ray lines (assuming it remains optically
thin).  Ref.~\cite{sk08}, on the other hand, assumed that the disk gas
is optically thick, and that the shocks are promptly dissipated in the
disk plane, and derived a characteristic infrared light--curve.  The
above effects arise from the change in the gravitational potential,
following the recoil, i.e. from the ``shaking'' of the disk.  A
different possibility is that once the BHs have merged, the torques
from the rotating quadrupolar potential cease to act on the gas
outside the inner cavity, allowing the gas to accrete onto the merged
binary on the viscous time-scale \cite{mp05}. An X--ray ``afterglow''
may then occur after a delay of $\sim$$7(1+z)(M/10^6{\rm
M_\odot})^{1.32}$ years. This could be detectable in {\it LISA}
follow-up observations, or perhaps without a {\it LISA} trigger for
very massive BHs at high $z$, where the time--scale is long.

\section*{Acknowledgments}

ZH thanks the organizers of the conference. The work described here
was supported by NASA (grant NNX08AH35G), and by the Pol\'anyi Program
of the Hungarian National Office for Research and Technology (NKTH).


\section*{References}
\begin{thebibliography}{9}


\bibitem{an02} P. J. Armitage, \& P. Natarajan, \textit{Astrophys.\ J.} \textbf{567}, L9 (2002).

\bibitem{al94} P. Artymowicz, \& S. H. Lubow, \textit{Astrophys.\ J.} \textbf{421}, 651 (1994).

\bibitem{al96} P. Artymowicz, \& S. H. Lubow, \textit{Astrophys.\ J. Lett.} \textbf{467}, L77 (1996).

\bibitem{bh92} J. E. Barnes, \& L. Hernquist, \textit{Ann. Rev. Ast. \& Astrophys.} \textbf{30}, 705 (1992).

\bibitem{campanelli07} M. Campanelli, et al., \textit{Phys. Rev. Lett.} \textbf{98}, 231102 (2007).

\bibitem{lia08} L. Corrales, A. MacFadyen, \& Z. Haiman, \textit{Astrophys.\ J.}, to be submitted (2008).

\bibitem{cuadra08} J. Cuadra, P. J. Armitage, R. D. Alexander, \& M. C. Begelman, \textit{Mon. Not. Royal Ast. Soc.}, submitted; arXiv:0809.0311

\bibitem{dm07} C. Deffayet, \& K. Menou, \textit{Astrophys.\ J.} \textbf{668}, L143 (2007).

\bibitem{elcm04a} A. Escala, R. B. Larson, P. S. Coppi, \& D. Mardones, \textit{Astrophys.\ J.} \textbf{607}, 765 (2004).

\bibitem{periodic} Z. Haiman, B. Kocsis, \& K. Menou, \textit{Astrophys.\ J.}, submitted; arXiv:0807.4697

\bibitem{hh05} D. E. Holz, \& S. A. Hughes, \textit{Astrophys.\ J.} \textbf{629}, 15 (2005).

\bibitem{ivanov99} P. B. Ivanov, J. C. B. Papaloizou, \& A. G. Polnarev, \textit{Mon. Not. Royal Ast. Soc.} \textbf{307}, 79 (1999).

\bibitem{koc06} B. Kocsis, Z. Frei, Z. Haiman, \& K. Menou, \textit{Astrophys.\ J.} \textbf{637}, 27 (2006).

\bibitem{paper1} B. Kocsis, Z. Haiman, K. Menou, \& Z. Frei, \textit{Phys.\ Rev.\ D} \textbf{76}, 022003 (2007).

\bibitem{paper2} B. Kocsis, Z. Haiman, \& K. Menou, \textit{Astrophys.\ J.} \textbf{684}, 870 (2008). 

\bibitem{kl08} B. Kocsis, \& A. Loeb, \textit{Phys. Rev. Lett.} \textbf{101}, 041101 (2008).

\bibitem{lh06} R. N. Lang, \& S. A. Hughes, \textit{Phys.\ Rev.\ D} \textbf{74}, 122001 (2006).

\bibitem{lh07} R. N. Lang, \& S. A. Hughes,  \textit{Astrophys.\ J.} \textbf{677}, 1184 (2008). 

\bibitem{lippai} Z. Lippai, Z. Frei, \& Z. Haiman, \textit{Astrophys.\ J. Lett.} \textbf{676}, L5 (2008). 

\bibitem{mm08} A. MacFadyen, \& M. Milosavljevi\'c, \textit{Astrophys.\ J.} \textbf{672}, 83 (2008). 

\bibitem{mad99} P. Madau, ``Cosmic Star Formation History and the
  Brightness of the Night Sky'', in {\it From Extrasolar Planets to
  Cosmology: The VLT Opening Symposium}, edited by J. Bergeron and
  A. Renzini (Springer-Verlag, 2000), p.\ 52.; arXiv:astro-ph/9907268


\bibitem{mefa01} F. Melia, \& H. Falcke, \textit{Ann. Rev. Ast. \& Astrophys.} \textbf{39}, 309 (2001).

\bibitem{mp05} M. Milosavljevic, \& E. S. Phinney, \textit{Astrophys.\ J.} \textbf{622}, L93 (2005).

\bibitem{sb08} G. A. Shields, \& E. W. Bonning \textit{Astrophys.\ J.} \textbf{682}, 758 (2008). 

\bibitem{sk08} J. D. Schnittman, \& J. H. Krolik, \textit{Astrophys.\ J.} \textbf{684}, 835 (2008). 

\bibitem{sch86} B. F. Schutz, \textit{Nature} \textbf{323}, 310 (1986).

\bibitem{htpulsar} J. H. Taylor, \&  J. M. Weisberg, \textit{Astrophys.\ J.} \textbf{253}, 908 (1992).

\bibitem{vec04} A. Vecchio, \textit{Phys.\ Rev.\ D} \textbf{70}, 042001 (2004).

\end{thebibliography}
\end{document}